\documentclass[
twocolumn,
]{ceurart}

\usepackage{booktabs}
\usepackage{colortbl}
\usepackage{graphicx,lipsum,afterpage,subcaption}
\usepackage{enumitem}
\usepackage{xspace}
\usepackage{float}
\usepackage{tabularx}

\newcommand{\LIME}{LIME\xspace}
\newcommand{\LIMERS}{LIME-RS\xspace}

\colorlet{tableheadcolor}{gray!25} 
\newcommand{\headcol}{\rowcolor{tableheadcolor}} %
\colorlet{tablerowcolor}{gray!10} 
\newcommand{\rowcol}{\rowcolor{tablerowcolor}} %
\newcommand{\topline}{\arrayrulecolor{black}\specialrule{0.1em}{\abovetopsep}{0pt}%
            \arrayrulecolor{tableheadcolor}\specialrule{\belowrulesep}{0pt}{0pt}%
            \arrayrulecolor{black}}
\newcommand{\midline}{\arrayrulecolor{tableheadcolor}\specialrule{\aboverulesep}{0pt}{0pt}%
            \arrayrulecolor{black}\specialrule{\lightrulewidth}{0pt}{0pt}%
            \arrayrulecolor{white}\specialrule{\belowrulesep}{0pt}{0pt}%
            \arrayrulecolor{black}}

\newcommand{\bottomlinec}{\arrayrulecolor{tablerowcolor}\specialrule{\aboverulesep}{0pt}{0pt}%
            \arrayrulecolor{black}\specialrule{\heavyrulewidth}{0pt}{\belowbottomsep}}%

\begin{document}

\copyrightyear{2021}
\copyrightclause{Copyright for this paper by its authors. Use permitted under Creative Commons License Attribution 4.0 International (CC BY 4.0).}

\conference{3rd Edition of Knowledge-aware and Conversational Recommender Systems (KaRS) \& 5th Edition of Recommendation in Complex Environments (ComplexRec) Joint Workshop @ RecSys 2021, September 27--1 October 2021, Amsterdam, Netherlands}

\title{Adherence and Constancy in LIME-RS Explanations for Recommendation}

 \author[1]{Vito Walter Anelli}[
]

\author[2]{Alejandro Bellogín}[
]

 \author[1]{Tommaso {Di Noia}}[
]

 \author[3]{Francesco Maria Donini}[
]

\author[1]{Vincenzo Paparella}[
]

\author[1]{Claudio Pomo}[
    email=claudio.pomo@poliba.it
]

\address[1]{Politecnico di Bari, via Orabona, 4, 70125 Bari, Italy}
\address[2]{Universidad Autónoma de Madrid, Ciudad Universitaria de Cantoblanco, 28049 Madrid, Spain}
\address[3]{Università degli Studi della Tuscia, via Santa Maria in Gradi, 4, 01100 Viterbo, Italy}

\begin{abstract}
  Explainable Recommendation 
  has attracted a lot of attention due to a renewed interest in explainable artificial intelligence. In particular, post-hoc approaches have proved to be the most easily applicable ones to increasingly complex recommendation models, which are then treated as black boxes.
The most recent literature has shown that for post-hoc explanations based on local surrogate models, there are problems related to the robustness of the approach itself. This consideration becomes even more relevant in human-related tasks like recommendation. 
  The explanation also has the arduous task of enhancing increasingly relevant aspects of user experience such as transparency or trustworthiness. This paper aims to show how the characteristics of a classical post-hoc model based on surrogates is strongly model-dependent and does not prove to be accountable for the explanations generated.
\end{abstract}


\maketitle

\section{Introduction}\label{sec:intro}
The explanation of a recommendation list plays an increasingly important role in the interaction of a user with a recommender system: the pervasiveness of economic interest and the inscrutability of most Artificial Intelligence systems make users ask for some form of accountability in the behavior of the systems they interact with. 
Given the explanation that a system can provide to a user we identify at least two characteristics that the explanation part should enforce~\cite{DBLP:journals/ai/Miller19, DBLP:reference/sp/TintarevM15, DBLP:journals/ijmms/GedikliJG14}:
\begin{itemize}[leftmargin=*]
    \item \emph{Adherence} to reality: the explanation should mention only features that really pertain to the recommended item. For instance, if the system recommends the movie ``Titanic'', it should not explain this recommendation by saying ``because it is a War Movie'' since it is by no means an adherent description of that movie;
    
    \item \emph{Constancy} in the behavior: when the explanation is generated based on some sample, and such a sample is drawn with a probability distribution, the entire process should not exhibit a random behavior to the user. For instance, if the explanation for recommending the movie ``The Matrix'' to the same user is first ``because it is a Dystopian Science Fiction'', and then ``because it is an Acrobatic Duels Movie'', for the same user, this behavior would be perceived as nondeterministic, 
    and thus reducing its trustworthiness.
\end{itemize}

Among several ways of generating explanations, we study here the application of \LIME~\cite{DBLP:conf/kdd/Ribeiro0G16} to the recommendation process. \LIME is an algorithm that can explain the predictions of any
classifier or regressor in a faithful way, by approximating it locally with an interpretable model. \LIME belongs to the category of post-hoc algorithms and it sees the prediction system as a black box by ignoring its underlying operations and algorithms. 
Since we can consider the recommendation task as a particular Machine Learning task, the \LIME  approach can also be applied to recommendation.
\LIMERS~\cite{DBLP:conf/sac/NobregaM19} is an adaptation of the general algorithm to the recommendation task and can be considered in all respects as a black-box explainer. This means that it generates an explanation by drawing a huge number of (random) calls to the system, collecting the answers, building a model of behavior of the system, and then constructing the explanation for the particular recommended item.
While the fact of adopting a black-box approach lets \LIMERS to be applicable for every recommender system, the way of building  a model by drawing a huge random sample of system behaviors makes it lose both adherence and constancy, as our experiments show later on this paper.
This suggests that the direct application of \LIMERS to recommender systems is not advisable, and that further research is needed to assess the usefulness of \LIMERS in explaining recommendations.




The paper is organized as follows: Section~\ref{sec:related} reviews the state of the art on explanation in recommendation; Section~\ref{sec:background} details LIME to make the paper self-contained. Section~\ref{sec:experiments} shows the results of experiments with two mainstream recommendation models: Attribute Item-kNN and Vector Space Model. We discuss the outcomes of the experiments in Section~\ref{sec:discussion}, and conclude with Section~\ref{sec:conclusion}.

\section{Related Work}\label{sec:related}
In recent years, the theme of Explanation in Artificial Intelligence has come to the foreground, capturing the attention not only of the Machine Learning and related communities -- that deal more specifically with the algorithmic part -- but also of fields closer to Social Sciences, such as Sociology or Cognitivism, which look with great interest to this area of research~\cite{DBLP:journals/ai/Miller19}. 
The growing interest in this area is also dictated by new regulations of both Europe~\cite{wachter2017counterfactual} and US~\cite{DBLP:conf/kbse/ChakrabortyPM20} with respect to sensitive issues in the field of personal data processing, and legal responsibility.
This trend has also touched the research field of recommender systems~\cite{DBLP:journals/ftir/ZhangC20, DBLP:conf/semweb/AnelliNSRT19, polleti2020explanations, DBLP:conf/ijcai/PanLLZ20}. However, topics such as explanation are by no means new to this field. In fact, we can date back to 2014 the introduction of the term ``explainable recommendation''~\cite{DBLP:conf/sigir/ZhangL0ZLM14}, although the need to provide an explanation that accompanies the recommendation is a need that emerged as early as 1999 by~\citet{DBLP:conf/sigecom/SchaferKR99}, when people began trying to explain a recommendation with other similar items familiar to the user who received that recommendation.

Catalyzation of interest around the topic of explanation of recommendations coincides also with the awareness achieved in considering metrics beyond accuracy as fundamental in evaluating a recommendation system~\cite{DBLP:conf/chi/McNeeRK06, DBLP:conf/sigir/Vargas14}. Indeed, all of the well-known metrics of novelty, diversity, and serendipity are intended to improve the user experience, and in this respect, a key role is played by explanation~\cite{DBLP:journals/ijmms/GedikliJG14, DBLP:conf/icde/TintarevM07}. ``Why are you recommending that?"---this is the question that usually accompanies the user when a suggestion is provided. \citet{DBLP:reference/sp/TintarevM15} detailed in a scrupulous way the aspects involved in the process of explanation when we talk about recommendation. They identified 7 aspects: user's trust, satisfaction, persuasiveness, efficiency, effectiveness, scrutability, and transparency.

This is the starting point to define Explainable Recommendation as a task that aims to provide suggestions to the users and make them aware of the recommendation process, explaining also why that specific object has been suggested. \citet{DBLP:journals/ijmms/GedikliJG14} evaluated different types of explanations and drew a set of guidelines to decide what the best explanation that should equip a recommendation system is. This is due to the fact that popular recommendation systems are based on Matrix Factorization (MF)~\cite{DBLP:journals/computer/KorenBV09}; for this type of model, trying to provide an explanation opens the way to new challenges~\cite{DBLP:journals/ai/Miller19, DBLP:conf/recsys/TsukudaG19, DBLP:conf/sigir/ChenCXZ0QZ19, DBLP:conf/caise/CornacchiaDNPR21}.

There are two different approaches to address this type of issue.
\begin{itemize}
    \item On the one hand, the model-intrinsic explanation strategy aims to create a user-friendly recommendation model or encapsulates an explaining mechanism. However, as~\citet{DBLP:journals/cacm/Lipton18} points out, this strategy will weigh in on the trade-off between the transparency and accuracy of the model. Indeed, if the goal becomes to justify recommendations, the purpose of the system is no longer to provide only personalized recommendations, resulting in a distortion of the recommendation process.
    
    \item On the other hand, we have a model-agnostic~\cite{DBLP:conf/icdm/WangCYWW018} approach, also known as post-hoc~\cite{DBLP:conf/kdd/PeakeW18}, which does not require to intervene on the internal mechanisms of the recommendation model and therefore does not affect its performance in terms of accuracy.
\end{itemize}
Most recommendation algorithms take an MF-approach, and thus the entire recommendation process is based on the interaction of latent factors that bring out the level of liking for an item with respect to a user. Many post-hoc explanation methods have been proposed for precisely these types of recommendation models. It seems evident that the most difficult challenge for this type of approach lies in making these latent factors explicit and understandable for the user~\cite{DBLP:conf/semweb/AnelliNSRT19}. \citet{DBLP:conf/kdd/PeakeW18} generate an explanation by exploiting the association rules between features; \citet{DBLP:conf/sigir/TaoJWW19} in their work, find benefit from regression trees to drive  learning, and then explain the latent space; instead, \citet{DBLP:conf/aaai/GaoWW019} try a deep model based on attention mechanisms to make relevant features emerge. Along the same lines are~\citet{DBLP:conf/ijcai/PanLLZ20}, who present a feature mapping approach that maps the uninterpretable general features onto the interpretable aspect features. Among other approaches to consider, \cite{DBLP:conf/sigir/ZhangL0ZLM14} proposes an explicit factor model that builds a mapping between the interpretable features and the latent space. On the same line we also find the work by~\citet{DBLP:conf/ijcai/FuscoVVWS19}. In their work, they provide an approach to identify, in a neural model, which features contribute most to the recommendation. However, these post-hoc explanation approaches turn out to be built for very specific models.  Purely model-agnostic approaches include the recent work of~\citet{DBLP:conf/iclr/TsangCLFZL20}, who present GLIDER, an approach to estimate interactions between features rather than on the significance of features as in the original \LIME~\cite{DBLP:conf/kdd/Ribeiro0G16} algorithm. This type of solution is constructed regardless of the recommendation model.

Our paper focuses on the operation of \LIME, a model-agnostic method for a surrogate-based local explanation. When a user-item pair is provided, this model returns as an outcome of the explanation a set of feature weights, for any recommender system.
However, the recommendation task is very specific, so there is a version called \LIMERS~\cite{DBLP:conf/sac/NobregaM19} that applies the explanation model technique to the recommendation domain. In this way, any recommender is seen as a black box, so \LIMERS plays the role of a model-agnostic explainer whose result is a set of interpretable features and their relative importance.

The goal of \LIMERS is to exploit the predictive power of the recommendation (black box) model to generate an explanation about the suggestion of a particular item for a user. In this respect, it exploits a neighborhood drawn according to a generic distribution compared to the candidate item for the explanation. It seems obvious that the choice of the neighborhood plays a crucial role within the process of explanation generation by \LIMERS. We can compare this sample extraction action to a perturbation of the user-item pair we are using to generate the explanation. In the case of \LIMERS this perturbation must generate consistent samples with respect to the source dataset. 
We see that this choice represents a critical issue for all the post-hoc models which base their expressiveness on the locality of the instance to explain.

This trend is confirmed in several papers addressing this issue of surrogate-based explanation systems such as LIME and SHAP~\cite{DBLP:journals/kais/StrumbeljK14}. In two recent papers,~\citet{DBLP:conf/nips/Alvarez-MelisJ18}  have shown how the explanations generated with LIME are not very robust: their contribution aims to bring out how small variations or perturbations in the input data cause significant variations in the explanation of that specific input~\cite{DBLP:journals/corr/abs-1806-08049}. In their paper, a new strategy is introduced to strengthen these methods by exploiting local Lipschitz continuity. By deeply investigating this drawback, they introduced self-explaining models in stages, progressively generalizing linear classifiers to complex yet architecturally explicit models.

\citet{DBLP:journals/corr/abs-2006-12302} also explored this issue by turning their gaze to different types of sampling to make the result of an explanation generated through LIME more robust. In particular, in their work, they introduce the possibility of generating realistic samples produced with a Generative Adversarial Network. Finally,~\citet{DBLP:conf/aies/SlackHJSL20} adopt a similar solution in order to control the perturbation generating neighborhood data points by attempting to mitigate the generation of unreliable explanations while maintaining a stable black-box model of prediction.

\section{Background Technology}\label{sec:background}
From a formal point of view, we can define a LIME-generated explanation for a generic instance $x \in \mathcal{X}$ produced by a model $f$ as:

\begin{equation}\label{lime}
\xi(x)=\underset{e \in E}{\operatorname{argmin}} \mathcal{L}\left(f, e, \pi_{x}\right)+\Omega(e)
\end{equation}
where $\mathcal{L}$ represents the fidelity of the surrogate model to the original $f$, and $e$ represents a particular instance of the class $E$ of all possible explainable models. Among all the possible models, the one most frequently chosen is based on a linear prediction. In this case, an explanation refers to the weights of the most important interpretable features, which, when combined, minimize the divergence from the black-box model. The function $ \pi_{x}$ measures the distance between the instance to be explained $x \in \mathcal{X}$, and the samples $x' \in \mathcal{X}$ extracted from the training set to train the model $e$. Finally, $\Omega(e)$ represents the complexity of the explanation model. 

Two pieces of evidence make the application of LIME possible: (i) the existence of a feature space $\mathcal{Z}$ on which to train the surrogate model of $f$, (ii) and the presence of a surjective function that maps the space mentioned above ($\mathcal{Z}$) to the original space of instances ($\mathcal{X}$).
Going into more detail, we consider the fidelity function $\mathcal{L}$ as the mean square deviation between the prediction for a generic instance $x' \in \mathcal{X}$ of the black-box model and that generated for the counterpart $z' \in \mathcal{Z}$ by the surrogate model. 
Starting from these considerations we can express $\mathcal{L}$ with the following formula:

\begin{equation}\label{limeloss}
\mathcal{L}\left(f, e, \pi_{x}\right)=\sum_{x' \in \mathcal{X}, z' \in \mathcal{Z}} \pi_{x}\left(x'\right)\cdot\left(f\left(x'\right)-e\left(z'\right)\right)^{2}
\end{equation}

In the formula above $\pi_{x}$ plays a fundamental role as it expresses the distance between the instance to be explained and the sampled instance used to build the surrogate model. From a generic perspective, we can express this function as a kernel function like $\pi_{x}=exp(-D(x,x')^{2}/\sigma^{2})$, where $D$ is any measure of distance. 

The full impact of this distance is captured when the fidelity function also considers the transformation of the surrogate sample in the original space. As mentioned earlier, we consider a surjective function $p$ that maps the original space into the feature space $p: \mathcal{X} \rightarrow \mathcal{Z}$. We can also consider the function that allows us to move in the opposite direction $p^{-1}: \mathcal{X} \rightarrow \mathcal{Z}$. At this point, Equation ~\eqref{limeloss} becomes:

\begin{equation}
\resizebox{.9\hsize}{!}{$
\mathcal{L}\left(f, e, \pi_{x}, p\right)=\sum_{z' \in \mathcal{Z}} \pi_{x}\left(p^{-1}\left(z'\right)\right)\cdot\left(f\left(p^{-1}\left(z'\right)\right)-e\left(z'\right)\right)^{2}$}
\end{equation}

From this last equation, we can grasp the criticality of the surjective mapping function. Indeed, the neighborhood in $\mathcal{Z}$-space cannot be guaranteed with the transformation in $\mathcal{X}$-space. Thus, some samples selected to train the surrogate model could not satisfy the neighborhood criterion for which they were chosen. 

We must therefore stress on the centrality of the sampling function: how do we extract the neighborhood of our instance to be explained?
If we look at the application of \LIME to the recommendation domain, we can compare this sampling action to a local perturbation around our instance $x$; however, this perturbation aims to generate $n$ samples $x'$, which might contain inconsistencies: as an example, suppose we want to explain \textit{James}'s feeling about the movie \textit{The Matrix}. The original triple of the instance to be explained associates to the user-item pair the genre of the movie (representing the explainable space) and in this case it is of the type $\langle James, The Matrix, Sci\text{-}Fi \rangle$. A perturbation around this instance could generate inconsistencies of the type $\langle James, The Matrix, Western\rangle$. For this reason, in \LIMERS the perturbation considers only real and not synthetic data. This choice is dictated by the avoidance of the out-of-sample (OOS) process phenomenon. Closely related to this problem predicted by OOS is that the interpretation examples selected in \LIMERS represent the ability to capture the locality through disturbance mechanisms effectively. One of the disadvantages of \LIME-like methods is that they sometimes fail to estimate an appropriate local replacement model but instead generate a model that focuses on explaining the examples and is also affected by more general trends in the data.

This issue is central to our work, and it involves two aspects: (i) the first one concerns the sampling function of the samples precisely. In the \LIMERS implementation, this function is driven by the popularity distribution of the items within the dataset. (ii) The second critical issue concerns the model's ability to wittily discriminate the user's taste from the neighborhood extracted to build the surrogate model. A model that squashes too much on bias or is inaccurate cannot bring out the peculiarities of user taste that are critical in building the explainable model which are, in turn, useful in generating the explanation for the instance of interest.

These observations dictate the two research questions that motivated our work: 
\begin{enumerate}
    \item[\textbf{RQ1}] Can we consider the surrogate-based model on which \LIMERS is built to generate always the same explanations, or does the extraction of a different neighborhood severely impact the system's constancy?
    \item[\textbf{RQ2}] Are \LIMERS explanations adherent to item content, despite the fact that the sampling function is uncritical and based only on popularity?
\end{enumerate}

\section{Experiments}\label{sec:experiments}
This section is devoted to illustrating how the experimental campaign was conducted. The datasets used for this phase of experimentation are \textit{Movielens 1M}~\cite{DBLP:journals/tiis/HarperK16}, \textit{Movielens Small}~\cite{DBLP:journals/tiis/HarperK16}, and \textit{Yahoo! Movies\footnote{R4 - Yahoo! Movies User Ratings and Descriptive Content Information, v.1.0 \url{http://webscope.sandbox.
yahoo.com/}.}}. Their characteristics are shown in Table~\ref{dataset}.

\begin{table}[htbp]
\caption{Characteristics of the datasets involved in the experiments.}
\resizebox{\linewidth}{!}{\begin{tabular}{lrrrr}
\topline
 \headcol& \multicolumn{1}{l}{Users} & \multicolumn{1}{l}{Items} & \multicolumn{1}{l}{Transactions} & \multicolumn{1}{l}{Sparsity} \\ \midline
Movielens 1M & 6040 & 3675 & 797758 & 0,9640 \\
\rowcol Movielens Small & 610 & 8990 & 80419 & 0,9853\\ 
Yahoo! Movies & 7636 & 8429 & 160604 & 0,9975 \\ \bottomlinec
\end{tabular}}
\label{dataset}
\end{table}

As for the choice of the models to be used in this work is concerned, we selected two well-known recommendation models that are able to exploit the information content of the items to produce a recommendation: Attribute Item kNN (Att-Item-kNN) and Vector Space Model (VSM). The two chosen models represent the simplest solution to address the recommendation problem by exploiting the content associated with the items in the catalog.

Att-Item-kNN exploits the characteristics of neighborhood-based models but expresses the representation of the items in terms of their content and, based on this representation, it computes a  similarity between users. Starting from this similarity and exploiting the collaborative contribution in terms of interactions
between users and items, Att-Item-kNN tries to estimate the level of liking of the items in the catalog. 
VSM represents both users and items in a new space to link users and items to the considered information content. Once obtained this new representation, with an appropriate function of similarity, VSM estimates which are the most appealing items for a specific user.
The implementation of both models are available in the ELLIOT~\cite{DBLP:conf/sigir/AnelliBFMMPDN21} evaluation framework. This benchmarking framework was used to select the best configuration for the two recommendation models by exploiting the corresponding configuration file\footnote{\url{https://tny.sh/basic_limers}}. 

Our experiments start by selecting the best configurations based on nDCG~\cite{DBLP:conf/recsys/AnelliNSPR19, DBLP:conf/kdd/KricheneR20} for the two models on the considered datasets. Then, we generate the top-10 list of recommendations for each user, and we take into account the first item $i_1$ on these lists for each user $u$. Finally, each recommendation pair $(u, i_1)$ is explained with \LIMERS. The explanation consists of a weighted vector $(g,w)_i$ where $g$ is the genre of the movies in the dataset -- \emph{i.e.}, the features -- and $w$ is the weight associated to $g$ by \LIMERS within the explanation. Then, this vector is sorted by descending weights. In this way, the genres of the movies which play a key role within the recommendation, as explained by \LIMERS, are highlighted at the first positions of the vector. These operations are then repeated $n=10$ times and changing the seed each time, as $10$ is likely to be a good choice to detect a general pattern in the behavior of \LIMERS. At this point, for each pair $(u,i_1)$, we have a group of $10$ explanations ordered by descending values of $w$, which we exploit to answer our two research questions.

\noindent \textbf{RQ1.}~\label{rq1}
Empirically, since in a real scenario of recommendation a too verbose explanation is not useful, we consider only the first five features in the sorted vector representing the explanation of each recommendation.
In order to verify the constancy of the behavior of \LIMERS, given a $(u,i_1)$ pair, 
we exploit the $n$ previously generated explanations for this pair. Then for $k=1,2,\ldots,5$, we define $G_k$ as the multiset of genres that appear in $k$-th position -- for instance, if ``Sci-Fi'' occurs in the first position of 7 explanations, then ``Sci-Fi'' occurs 7 times in the multiset $G_1$, and similarly for other genres and multisets.
Then, we compute the frequency of genres in each position as follows: given a position $k$, a genre $g$, and the number $n$ of generated explanations for a given pair $(u, i_1)$, the frequency $f_{g_{k}}$ of $g$ in $k$-th position is computed as:
\newcommand{\card}[1]{\ensuremath{\vert\vert #1 \vert\vert}}
\begin{equation}
    f_{g_{k}} = \frac{\card{\{g ~|~ g\in G_k\}}}{n}
\end{equation}
where $\card{\cdot}$ denotes the cardinality of a multiset.
Then, all this information is collected for each user in five lists --- one for each of the $k$ positions --- of pairs  $\langle g,f_{g_{k}} \rangle$ sorted by frequency. One can observe that the computed frequency is an estimation of the probability that a given genre is put in that position within the explanation generated by \LIMERS sorted by values. Hence, the pair $\langle g, \max \left( f_{g_{k}} \right) \rangle$ describes the genre with the highest frequency in the $k$-th position of the explanation for a pair $(u, i_1)$. Finally, it makes sense to compute the mean $\mu_k$ of the highest probability values in each position $k$ of the explanations for each pair $(u, i_1)$. Formally, by setting a position $k$, the mean $\mu_k$ is computed as:
\begin{equation}
    \mu_k= \frac{\sum_{j=1}^{|U|} \max \left( f_{g_{k}} \right) _j}{\vert U\vert}
\end{equation}
where $U$ is the set of users for whom it was possible to generate a recommendation for. Observing the value of $\mu_k$, we can state to what extent \LIMERS is constant in providing the explanations until the $k$-th 
feature: the higher the value of $\mu_k$, the higher the constancy of \LIMERS concerning the $k$-th feature.

\begin{table}[htbp]
\caption{Constancy of \protect{\LIMERS}. A value equal  to 0 means that the genre(s) provided by \protect{\LIMERS} in the first $k$ position(s) is always different (worst case: completely inconstant behavior); A value equal to 1 means that the genre(s) provided by \protect{\LIMERS} in the first $k$ position(s) is always the same (total constancy).}
\label{table:disp}
\resizebox{\linewidth}{!}{\begin{tabular}{llllll}
\topline
\headcol& \multicolumn{1}{c}{$\mu_1$} & \multicolumn{1}{c}{$\mu_2$} & \multicolumn{1}{c}{$\mu_3$} & \multicolumn{1}{c}{$\mu_4$} & \multicolumn{1}{c}{$\mu_5$} \\ \bottomlinec
\multicolumn{6}{c}{Att-Item-kNN}      

    \\ \bottomlinec
\rowcol Movielens 1M    & 0,9130 & 0,7822 & 0,6927 & 0,6288 & 0,5727 \\
Movielens Small & 0,8830 & 0,7426 & 0,6639 & 0,60459  & 0,5616 \\

\rowcol Yahoo! Movies            & \textbf{0,9230} & \textbf{0,8016} & \textbf{0,7232} & \textbf{0,6528} & \textbf{0,5830} \\ \bottomlinec
\multicolumn{6}{c}{VSM}                                         \\ \bottomlinec

\rowcol Movielens 1M    & 0,8929 & 0,7953 & 0,7729 & 0,7726 & 0,7801 \\
Movielens Small & 0,9464 & 0,8636 & 0,8343 & 0,8138  & 0,8049 \\

\rowcol Yahoo! Movies            & \textbf{0,9732} & \textbf{0,9209} & \textbf{0,8887} & \textbf{0,8884}  & \textbf{0,9056} \\ \bottomlinec
\end{tabular}}
\end{table}

By looking at Table~\ref{table:disp}, we can see that for Att-Item-kNN the \LIMERS explanation model is reliable as long as it considers at most three features in the weighted vector presented as an explanation of the recommendation. Extending the explanation to four features, we have a constancy that falls below 65\%, while arriving at an explanation with five features is more likely to run into explanations that exhibit an unacceptably random behavior.
On the other hand, we can see that for VSM the values are much more stable. In this case, we have a constancy that, regardless of the length of the weighted vector of the explanation, stabilizes on average around 80\%. An aspect emerges that will be discussed in detail later: \LIMERS is conditioned by the ability of the black-box model to discriminate the user's tastes locally.


\noindent \textbf{RQ2.}~\label{rq2}
With the aim of providing an answer about the adherence to reality of \LIMERS, we make a comparison between the genres claimed to explain a recommended item and its actual genres. Indeed, the explanations about an item should fit the list of genres the item is characterized by. This means that, in an ideal case, all highly weighted features within the explanation should match the genres of the item. From the results in Table~\ref{table:disp}, we notice that using Att-Item-kNN the constancy of \LIMERS reaches a low value after the third feature. Hence, it is a futile effort to go deeper in the study of the explanation. To this aim, we intersected 
each explanation limited to the set $E_k$ of its first $k$ genres with the set of genres $F_{i_1}$ characterizing the first  recommended item, for $k=1,2,3$. Upon completion of this operation for all the $n$ explanations generated for each $(u, i_1)$ pair, we computed the number of times we obtained an empty intersection of these sets, normalized by the total number of explanations $n \times |U|$, in order to understand to what extent an explanation is (not) adherent to the item. Formally, 
for a given value of $k$,
the value $adherence_k$ is computed as:
\begin{equation}
    adherence_k = \frac{\sum_{j=1}^{n \times |U|} \left[ \left( E_k \cap F_{i_1}\right)_j = \emptyset \right]}{n \times |U|}
\end{equation}
where $U$ is the set of users of the dataset for whom it was possible to generate a recommendation, $n$ is the number of generated explanations for each pair $(u, i_1)$, and by $\Sigma[\cdots]$ we mean that we sum 1 if the condition inside $[\cdots]$ is true, and 0 otherwise. One can note that $adherence_k \in [0,1]$, where a value of 1 indicates the worst case in which for none of the $n$ explanations under consideration at least one genre of the item is in the first $k$ features of the explanation. In contrast, the lower the value of $adherence_k$, the higher the adherence of \LIMERS. 

\begin{table}[htbp]
\caption{Adherence of \protect{\LIMERS}. For value equals to 1 no genre provided by \protect{\LIMERS} in the first $k$ real genres of the movie (worst case); For value equals to 0 at least one genre provided by \protect{\LIMERS} in the first $k$ genres
is always among the real genres of the movie.}
\label{table:check}
\resizebox{\linewidth}{!}{\begin{tabular}{llll}
\topline
\headcol& \multicolumn{1}{c}{$adherence_1$} & \multicolumn{1}{c}{$adherence_2$} & \multicolumn{1}{c}{$adherence_3$} \\ \bottomlinec
\multicolumn{4}{c}{Att-Item-kNN}                                                                                                                                         \\ \bottomlinec
\rowcol Movielens 1M    & \multicolumn{1}{c}{0,2774}                                              & \multicolumn{1}{c}{0,1105}                                               & \multicolumn{1}{c}{0,0488}                                              \\
Movielens Small & \multicolumn{1}{c}{\textbf{0,2364}}                                              & \multicolumn{1}{c}{\textbf{0,0651}}                                              & \multicolumn{1}{c}{\textbf{0,0180}}                                         \\

\rowcol Yahoo! Movies            & \multicolumn{1}{c}{0,3597}                                              & \multicolumn{1}{c}{0,1202}                                               & \multicolumn{1}{c}{0,0476}                                               \\ \bottomlinec
\multicolumn{4}{c}{VSM}                                                                                                                                                          \\ \bottomlinec

\rowcol Movielens 1M    & \multicolumn{1}{c}{0,5357}                                              & \multicolumn{1}{c}{0,2539}                                              & \multicolumn{1}{c}{0,1088}                                              \\
Movielens Small & \multicolumn{1}{c}{0,4384}                                              & \multicolumn{1}{c}{0,1674}                                              & \multicolumn{1}{c}{0,0403}                                              \\

\rowcol Yahoo! Movies            & \multicolumn{1}{c}{\textbf{0,1013}}                                              & \multicolumn{1}{c}{\textbf{0,01348}}                                              & \multicolumn{1}{c}{\textbf{0,0021}}                                              \\ \bottomlinec
\end{tabular}}
\end{table}

Observing the results from Table \ref{table:check}, Att-Item-KNN performs well in terms of adherence since, in approximately 75\% of cases, even considering only the main feature of the explanation, it falls into the set of the item genres, as for Movielens dataset family. This performance is a 10\% lower for Yahoo! Movies. In contrast with this result, VSM shows poor performances on both dataset of the Movielens family, by failing half the time about Movielens 1M as regards adherence. A surprising result is achieved  for Yahoo! Movies dataset because, enlarging the study to the first three features among the explanation, the error is almost completely absent. The reasons we found to explain this difference in the performances concern the characteristics and the quality of the dataset, as we highlight later on.

\section{Discussion}\label{sec:discussion}

This work investigates how well a \emph{post-hoc} approach based on local surrogates -- such as the \LIMERS algorithm -- explains a recommendation.
Instead of studying the impact of explanations on users (that is a well-studied topic in the literature and is beyond our scope), we focus on objective evidences that could emerge. In this respect, we have designed specific experiments, which introduced two different metrics, to evaluate adherence and constancy for this kind of algorithms. For instance, Table~\ref{table:disp} shows a different behavior for Att-Item-kNN and VSM. On the one hand, Att-Item-kNN seems to guarantee a good constancy in explanations up to the third feature. 
This suggests that an explanation that exploits the first three features of the list produced by \LIMERS could be barely considered as reliable (i.e., reaching a constancy of $0.69$ on Movielens 1M). On the other hand, VSM exhibits a much more "stable" behavior, demonstrating in all cases (except for the first feature with Movielens 1M) better performance than Att-Item-kNN in terms of constancy, with peaks up to 97\%.
A straightforward consequence of these observations could be analyzed in terms of confidence or probability.
If the constancy steadily decreases, it means that the probability that \LIMERS suggests the same explanatory feature decreases. In practical terms, we could say that \LIMERS is less confident about its explanation. In fact, this is the behavior of Att-Item-kNN.
Conversely, VSM shows high values of constancy, resulting in a more "deterministic" behavior. With VSM, \LIMERS is more confident of its explanations. This could increase user's trustworthiness, since \LIMERS behavior is more reliable.

However, these results could also be interpreted together with the ones from Table~\ref{table:check}. They show how often at least one feature -- out of $k$ features provided by \LIMERS -- adheres to the features that describe the item being explained. In other words, they measure the probability that \LIMERS succeeds in reconstructing at least one feature of a specific item. Combining the results of Table~\ref{table:disp} and those of Table~\ref{table:check},  Att-Item-kNN, as already mentioned, shows good performance regarding adherence and identifies 3~times out of~4 the first fundamental feature of the explanation among those present in the set of features originally associated with the item. As expected, if the number $k$ of \LIMERS-reconstructed features increases, the number of times such a set has a nonempty intersection (with the features belonging to the item) -- \emph{i.e.,} adherence -- increases. It could be noted that Att-Item-kNN on Yahoo! Movies shows the worst behavior in terms of adherence.
VSM shows a different behavior. Despite the excellent performance regarding constancy, it could be observed that on both Movielens datasets, the performance in terms of adherence is poor, and worse for Movielens 1M than for Movielens Small. Surprisingly, on Yahoo! Movies, VSM performs much better, and the errors are almost negligible. 

\noindent \textbf{The difference between the two models} could be due to many reasons. In the following we analyze possible relations between such behaviors and two of them: \textit{popularity bias in the dataset} and \textit{characteristics of side information}. On the one hand, if the dataset is affected by popularity bias, it would be a well-studied cause of confusion for \LIMERS. 
On the other hand, the characteristics of the side information associated with the datasets could dramatically influence the performance of the two recommendation models.
To assess these hypotheses, we have evaluated (see Table~\ref{table:perf}) the recommendation lists produced by Att-Item-kNN and VSM considering nDCG, Hit Rate (HR), Mean Average Precision (MAP), and Mean Reciprocal Rank (MRR).
\begin{table}[htbp]
\caption{Results of the experiments on the models involved in the experiments. Models are optimized according to the value of nDCG.}
\label{table:perf}
\resizebox{\linewidth}{!}{\begin{tabular}{lcccccc}
\topline
\headcol model      & nDCG   & Recall & HR     & Precision & MAP    & MRR    \\ \bottomlinec
\multicolumn{7}{c}{Movielens 1m}                                   \\ \bottomlinec

\rowcol Random     & 0,0051 & 0,0028 & 0,0869 & 0,0098    & 0,0094 & 0,0264 \\
MostPop    & 0,0845 & 0,0379 & 0,4548 & 0,104     & 0,115  & 0,2205 \\

\rowcol Att-Item-kNN & 0,0229 & 0,0165 & 0,2425 & 0,0383    & 0,0387 & 0,0888 \\
VSM        & 0,0173 & 0,0109 & 0,2106 & 0,0292    & 0,0306 & 0,0741 \\ \bottomlinec

\multicolumn{7}{c}{Movielens Small}        \\ \bottomlinec

\rowcol Random     & 0,0030  & 0,0013 & 0,0492 & 0,0049    & 0,0068 & 0,0205 \\

MostPop    & 0,0715 & 0,0389 & 0,3902 & 0,0748    & 0,0912 & 0,1961 \\

\rowcol Att-Item-kNN & 0,0124 & 0,0068 & 0,1459 & 0,0197    & 0,0191 & 0,0484 \\

VSM        & 0,0085  & 
0,0056  & 0,1000  & 0,0111     & 0,0123  & 0.0350  \\ \bottomlinec
\multicolumn{7}{c}{Yahoo! Movies}                                           \\ \bottomlinec

\rowcol Random     & 0,0005 & 0,0008 & 0,0051 & 0,0005    & 0,0005 & 0,0015 \\
MostPop    & 0,2188 & 0,2589 & 0,596  & 0,1067    & 0,1501 & 0,3447 \\

\rowcol Att-Item-kNN & 0,0215 & 0,0262 & 0,1198 & 0,0132    & 0,0155 & 0,0435 \\
VSM        & 0,0131 & 0,0171 & 0,0754 & 0,0081    & 0,0092 & 0,0261 \\ \bottomlinec
\end{tabular}}
\end{table}
Table~\ref{table:perf} shows that the chosen datasets are strongly affected by popularity bias. Indeed, MostPop is the best performing approach, and the two "personalized" models fail to produce accurate results.  
This triggers the second aspect that concerns the quality of the content.
The results suggest that the side information is not good enough to boost the recommendation systems in producing meaningful recommendations. In fact, the three datasets seem to have an informative content that is not adequate to generate appealing recommendations. We observe that, from an informative point of view, the Yahoo! Movies dataset is slightly more complete: 22 genres against the 18 genres available on Movielens. 
Although the VSM model does not show excellent performance, in combination with \LIMERS, it provides explanations that are very reliable in terms of constancy (see Table~\ref{table:disp}) and adherence (see Table~\ref{table:check}) to the actual content of the items being explained.

\noindent \textbf{From the designer perspective,} 
there is also a pragmatic way to look at the experimental results. Suppose a developer needs an off-the-shelf way of generating explanations for recommendations, and chooses \LIMERS to do that. Our results suggest that if the explainer employs a Movielens dataset with Att-Item-kNN model, then it is better to run the explainer several times.
Indeed, the first feature obtained for the explanation could change around 1 time every 5 trials (first column of Table~\ref{table:disp}), and once such a feature is obtained, it is better to check whether this feature is really among the ones describing the item, since 1 time out of 4 the feature can be wrong (first column of Table~\ref{table:check}).
Moreover, if the explainer employs the Yahoo! Movies dataset with VSM model, then probably there is no need to run the explainer twice, since its behavior is constant 97\% of the times, while the feature is wrong only 10\% of the times. However, the low performance of such a model is to be taken into account.
    

\section{Conclusion}\label{sec:conclusion}

In this paper we shed a first light on the effectiveness of \LIMERS as a black-box explanation model in a recommendation scenario. We propose two different measures to understand how reliable an explanation based on \LIMERS is: \emph{(i)} \emph{constancy} was used to 
assess the impact of the random sampling phase of \LIMERS on the provided explanation -- ideally the explanation should remain constant in spite of the sample used to obtain it; 
\emph{(ii)} \textit{adherence} was proposed to understand the reconstructive power of \LIMERS with respect to the features that belong to the item involved in the explanation -- ideally, \LIMERS should provide an explanation that always adheres to the actual features of the recommended item. 

To test both constancy and adherence, we trained and optimized two content-based recommendation models:  Attribute Item-kNN (Att-Item-kNN), and  a classical Vector Space Model. For each model, and for all datasets exploited in the study, we generated recommendation lists for all users. We exploited the first item of these top-10 lists to produce the explanations that were then the subject of our investigation.
It turned out that for models built with a large collaborative input such as Att-Item-kNN, \LIMERS produces fairly constant explanations up to a length of three features. Moreover, these explanations turn out to be adherent with respect to the item between 65\% and 75\% of the cases in which only the first feature of the weighted vector of explanations is considered.
VSM shows a different behavior where explanations are much more constant, but suffer a lot in terms of adherence, except for the Yahoo! Movies dataset for which the explanation model showed outstanding performance despite the poor ability of VSM to provide sound recommendations to users.

In our experiments, some evidence started to emerge highlighting that the adopted explanation model is conditioned not only by the accuracy of the black-box model it tries to explain but also by the quality of the side information used to train the model. The latter result deserves to be adequately investigated to search for a link at a higher level of detail.
We plan to apply our experiments also to other recommendation models, to 
see whether the problems with adherence and constancy that we found for the two tested models show up also in other situations. We will also investigate what impact structured knowledge has on this performance by exploiting models capable of leveraging this type of content. In addition, it would also be the case to try different reference domains with richer datasets of side information to understand what impact content quality has on this type of explainer.


\section*{Acknowledgments}
\small
The authors acknowledge partial support of PID2019-108965GB-I00, PON ARS01\_00876 BIO-D, Casa delle Tecnologie Emergenti della Città di Matera, PON ARS01\_00821 FLET4.0, PIA Servizi Locali 2.0, H2020 Passapartout - Grant n. 101016956, PIA ERP4.0, and IPZS-PRJ4\_IA\_NORMATIVO.

\bibliography{sample-ceur}

\begin{thebibliography}{36}
\expandafter\ifx\csname natexlab\endcsname\relax\def\natexlab#1{#1}\fi
\providecommand{\url}[1]{\texttt{#1}}
\providecommand{\href}[2]{#2}
\providecommand{\path}[1]{#1}
\providecommand{\DOIprefix}{doi:}
\providecommand{\ArXivprefix}{arXiv:}
\providecommand{\URLprefix}{URL: }
\providecommand{\Pubmedprefix}{pmid:}
\providecommand{\doi}[1]{\href{http://dx.doi.org/#1}{\path{#1}}}
\providecommand{\Pubmed}[1]{\href{pmid:#1}{\path{#1}}}
\providecommand{\bibinfo}[2]{#2}
\ifx\xfnm\relax \def\xfnm[#1]{\unskip,\space#1}\fi
\bibitem[{Miller(2019)}]{DBLP:journals/ai/Miller19}
\bibinfo{author}{T.~Miller},
\newblock \bibinfo{title}{Explanation in artificial intelligence: Insights from
  the social sciences},
\newblock \bibinfo{journal}{Artif. Intell.} \bibinfo{volume}{267}
  (\bibinfo{year}{2019}) \bibinfo{pages}{1--38}. \URLprefix
  \url{https://doi.org/10.1016/j.artint.2018.07.007}.
  \DOIprefix\doi{10.1016/j.artint.2018.07.007}.
\bibitem[{Tintarev and Masthoff(2015)}]{DBLP:reference/sp/TintarevM15}
\bibinfo{author}{N.~Tintarev}, \bibinfo{author}{J.~Masthoff},
\newblock \bibinfo{title}{Explaining recommendations: Design and evaluation},
\newblock in: \bibinfo{booktitle}{Recommender Systems Handbook},
  \bibinfo{publisher}{Springer}, \bibinfo{year}{2015}, pp.
  \bibinfo{pages}{353--382}.
\bibitem[{Gedikli et~al.(2014)Gedikli, Jannach, and
  Ge}]{DBLP:journals/ijmms/GedikliJG14}
\bibinfo{author}{F.~Gedikli}, \bibinfo{author}{D.~Jannach},
  \bibinfo{author}{M.~Ge},
\newblock \bibinfo{title}{How should {I} explain? {A} comparison of different
  explanation types for recommender systems},
\newblock \bibinfo{journal}{Int. J. Hum. Comput. Stud.} \bibinfo{volume}{72}
  (\bibinfo{year}{2014}) \bibinfo{pages}{367--382}. \URLprefix
  \url{https://doi.org/10.1016/j.ijhcs.2013.12.007}.
  \DOIprefix\doi{10.1016/j.ijhcs.2013.12.007}.
\bibitem[{Ribeiro et~al.(2016)Ribeiro, Singh, and
  Guestrin}]{DBLP:conf/kdd/Ribeiro0G16}
\bibinfo{author}{M.~T. Ribeiro}, \bibinfo{author}{S.~Singh},
  \bibinfo{author}{C.~Guestrin},
\newblock \bibinfo{title}{"why should {I} trust you?": Explaining the
  predictions of any classifier},
\newblock in: \bibinfo{editor}{B.~Krishnapuram}, \bibinfo{editor}{M.~Shah},
  \bibinfo{editor}{A.~J. Smola}, \bibinfo{editor}{C.~C. Aggarwal},
  \bibinfo{editor}{D.~Shen}, \bibinfo{editor}{R.~Rastogi} (Eds.),
  \bibinfo{booktitle}{Proceedings of the 22nd {ACM} {SIGKDD} International
  Conference on Knowledge Discovery and Data Mining, San Francisco, CA, USA,
  August 13-17, 2016}, \bibinfo{publisher}{{ACM}}, \bibinfo{year}{2016}, pp.
  \bibinfo{pages}{1135--1144}. \URLprefix
  \url{https://doi.org/10.1145/2939672.2939778}.
  \DOIprefix\doi{10.1145/2939672.2939778}.
\bibitem[{N{\'{o}}brega and Marinho(2019)}]{DBLP:conf/sac/NobregaM19}
\bibinfo{author}{C.~N{\'{o}}brega}, \bibinfo{author}{L.~B. Marinho},
\newblock \bibinfo{title}{Towards explaining recommendations through local
  surrogate models},
\newblock in: \bibinfo{editor}{C.~Hung}, \bibinfo{editor}{G.~A. Papadopoulos}
  (Eds.), \bibinfo{booktitle}{Proceedings of the 34th {ACM/SIGAPP} Symposium on
  Applied Computing, {SAC} 2019, Limassol, Cyprus, April 8-12, 2019},
  \bibinfo{publisher}{{ACM}}, \bibinfo{year}{2019}, pp.
  \bibinfo{pages}{1671--1678}. \URLprefix
  \url{https://doi.org/10.1145/3297280.3297443}.
  \DOIprefix\doi{10.1145/3297280.3297443}.
\bibitem[{Wachter et~al.(2017)Wachter, Mittelstadt, and
  Russell}]{wachter2017counterfactual}
\bibinfo{author}{S.~Wachter}, \bibinfo{author}{B.~Mittelstadt},
  \bibinfo{author}{C.~Russell},
\newblock \bibinfo{title}{Counterfactual explanations without opening the black
  box: Automated decisions and the gdpr},
\newblock \bibinfo{journal}{Harv. JL \& Tech.} \bibinfo{volume}{31}
  (\bibinfo{year}{2017}) \bibinfo{pages}{841}.
\bibitem[{Chakraborty et~al.(2020)Chakraborty, Peng, and
  Menzies}]{DBLP:conf/kbse/ChakrabortyPM20}
\bibinfo{author}{J.~Chakraborty}, \bibinfo{author}{K.~Peng},
  \bibinfo{author}{T.~Menzies},
\newblock \bibinfo{title}{Making fair {ML} software using trustworthy
  explanation},
\newblock in: \bibinfo{booktitle}{35th {IEEE/ACM} International Conference on
  Automated Software Engineering, {ASE} 2020, Melbourne, Australia, September
  21-25, 2020}, \bibinfo{publisher}{{IEEE}}, \bibinfo{year}{2020}, pp.
  \bibinfo{pages}{1229--1233}. \URLprefix
  \url{https://doi.org/10.1145/3324884.3418932}.
  \DOIprefix\doi{10.1145/3324884.3418932}.
\bibitem[{Zhang and Chen(2020)}]{DBLP:journals/ftir/ZhangC20}
\bibinfo{author}{Y.~Zhang}, \bibinfo{author}{X.~Chen},
\newblock \bibinfo{title}{Explainable recommendation: {A} survey and new
  perspectives},
\newblock \bibinfo{journal}{Found. Trends Inf. Retr.} \bibinfo{volume}{14}
  (\bibinfo{year}{2020}) \bibinfo{pages}{1--101}. \URLprefix
  \url{https://doi.org/10.1561/1500000066}. \DOIprefix\doi{10.1561/1500000066}.
\bibitem[{Anelli et~al.(2019)Anelli, Noia, Sciascio, Ragone, and
  Trotta}]{DBLP:conf/semweb/AnelliNSRT19}
\bibinfo{author}{V.~W. Anelli}, \bibinfo{author}{T.~D. Noia},
  \bibinfo{author}{E.~D. Sciascio}, \bibinfo{author}{A.~Ragone},
  \bibinfo{author}{J.~Trotta},
\newblock \bibinfo{title}{How to make latent factors interpretable by feeding
  factorization machines with knowledge graphs},
\newblock in: \bibinfo{editor}{C.~Ghidini}, \bibinfo{editor}{O.~Hartig},
  \bibinfo{editor}{M.~Maleshkova}, \bibinfo{editor}{V.~Sv{\'{a}}tek},
  \bibinfo{editor}{I.~F. Cruz}, \bibinfo{editor}{A.~Hogan},
  \bibinfo{editor}{J.~Song}, \bibinfo{editor}{M.~Lefran{\c{c}}ois},
  \bibinfo{editor}{F.~Gandon} (Eds.), \bibinfo{booktitle}{The Semantic Web -
  {ISWC} 2019 - 18th International Semantic Web Conference, Auckland, New
  Zealand, October 26-30, 2019, Proceedings, Part {I}}, volume
  \bibinfo{volume}{11778} of \textit{\bibinfo{series}{Lecture Notes in Computer
  Science}}, \bibinfo{publisher}{Springer}, \bibinfo{year}{2019}, pp.
  \bibinfo{pages}{38--56}. \URLprefix
  \url{https://doi.org/10.1007/978-3-030-30793-6\_3}.
  \DOIprefix\doi{10.1007/978-3-030-30793-6\_3}.
\bibitem[{Polleti et~al.(2020)Polleti, Munhoz, and
  Cozman}]{polleti2020explanations}
\bibinfo{author}{G.~P. Polleti}, \bibinfo{author}{H.~N. Munhoz},
  \bibinfo{author}{F.~G. Cozman},
\newblock \bibinfo{title}{Explanations within conversational recommendation
  systems: improving coverage through knowledge graph embedding},
\newblock in: \bibinfo{booktitle}{2020 AAAI Workshop on Interactive and
  Conversational Recommendation System. AAAI Press, New York City, New York,
  USA}, \bibinfo{year}{2020}.
\bibitem[{Pan et~al.(2020)Pan, Li, Li, and Zhu}]{DBLP:conf/ijcai/PanLLZ20}
\bibinfo{author}{D.~Pan}, \bibinfo{author}{X.~Li}, \bibinfo{author}{X.~Li},
  \bibinfo{author}{D.~Zhu},
\newblock \bibinfo{title}{Explainable recommendation via interpretable feature
  mapping and evaluation of explainability},
\newblock in: \bibinfo{editor}{C.~Bessiere} (Ed.),
  \bibinfo{booktitle}{Proceedings of the Twenty-Ninth International Joint
  Conference on Artificial Intelligence, {IJCAI} 2020},
  \bibinfo{publisher}{ijcai.org}, \bibinfo{year}{2020}, pp.
  \bibinfo{pages}{2690--2696}. \URLprefix
  \url{https://doi.org/10.24963/ijcai.2020/373}.
  \DOIprefix\doi{10.24963/ijcai.2020/373}.
\bibitem[{Zhang et~al.(2014)Zhang, Lai, Zhang, Zhang, Liu, and
  Ma}]{DBLP:conf/sigir/ZhangL0ZLM14}
\bibinfo{author}{Y.~Zhang}, \bibinfo{author}{G.~Lai},
  \bibinfo{author}{M.~Zhang}, \bibinfo{author}{Y.~Zhang},
  \bibinfo{author}{Y.~Liu}, \bibinfo{author}{S.~Ma},
\newblock \bibinfo{title}{Explicit factor models for explainable recommendation
  based on phrase-level sentiment analysis},
\newblock in: \bibinfo{editor}{S.~Geva}, \bibinfo{editor}{A.~Trotman},
  \bibinfo{editor}{P.~Bruza}, \bibinfo{editor}{C.~L.~A. Clarke},
  \bibinfo{editor}{K.~J{\"{a}}rvelin} (Eds.), \bibinfo{booktitle}{The 37th
  International {ACM} {SIGIR} Conference on Research and Development in
  Information Retrieval, {SIGIR} '14, Gold Coast , QLD, Australia - July 06 -
  11, 2014}, \bibinfo{publisher}{{ACM}}, \bibinfo{year}{2014}, pp.
  \bibinfo{pages}{83--92}. \URLprefix
  \url{https://doi.org/10.1145/2600428.2609579}.
  \DOIprefix\doi{10.1145/2600428.2609579}.
\bibitem[{Schafer et~al.(1999)Schafer, Konstan, and
  Riedl}]{DBLP:conf/sigecom/SchaferKR99}
\bibinfo{author}{J.~B. Schafer}, \bibinfo{author}{J.~A. Konstan},
  \bibinfo{author}{J.~Riedl},
\newblock \bibinfo{title}{Recommender systems in e-commerce},
\newblock in: \bibinfo{editor}{S.~I. Feldman}, \bibinfo{editor}{M.~P. Wellman}
  (Eds.), \bibinfo{booktitle}{Proceedings of the First {ACM} Conference on
  Electronic Commerce (EC-99), Denver, CO, USA, November 3-5, 1999},
  \bibinfo{publisher}{{ACM}}, \bibinfo{year}{1999}, pp.
  \bibinfo{pages}{158--166}. \URLprefix
  \url{https://doi.org/10.1145/336992.337035}.
  \DOIprefix\doi{10.1145/336992.337035}.
\bibitem[{McNee et~al.(2006)McNee, Riedl, and
  Konstan}]{DBLP:conf/chi/McNeeRK06}
\bibinfo{author}{S.~M. McNee}, \bibinfo{author}{J.~Riedl},
  \bibinfo{author}{J.~A. Konstan},
\newblock \bibinfo{title}{Being accurate is not enough: how accuracy metrics
  have hurt recommender systems},
\newblock in: \bibinfo{editor}{G.~M. Olson}, \bibinfo{editor}{R.~Jeffries}
  (Eds.), \bibinfo{booktitle}{Extended Abstracts Proceedings of the 2006
  Conference on Human Factors in Computing Systems, {CHI} 2006, Montr{\'{e}}al,
  Qu{\'{e}}bec, Canada, April 22-27, 2006}, \bibinfo{publisher}{{ACM}},
  \bibinfo{year}{2006}, pp. \bibinfo{pages}{1097--1101}. \URLprefix
  \url{https://doi.org/10.1145/1125451.1125659}.
  \DOIprefix\doi{10.1145/1125451.1125659}.
\bibitem[{Vargas(2014)}]{DBLP:conf/sigir/Vargas14}
\bibinfo{author}{S.~Vargas},
\newblock \bibinfo{title}{Novelty and diversity enhancement and evaluation in
  recommender systems and information retrieval},
\newblock in: \bibinfo{editor}{S.~Geva}, \bibinfo{editor}{A.~Trotman},
  \bibinfo{editor}{P.~Bruza}, \bibinfo{editor}{C.~L.~A. Clarke},
  \bibinfo{editor}{K.~J{\"{a}}rvelin} (Eds.), \bibinfo{booktitle}{The 37th
  International {ACM} {SIGIR} Conference on Research and Development in
  Information Retrieval, {SIGIR} '14, Gold Coast , QLD, Australia - July 06 -
  11, 2014}, \bibinfo{publisher}{{ACM}}, \bibinfo{year}{2014}, p.
  \bibinfo{pages}{1281}. \URLprefix
  \url{https://doi.org/10.1145/2600428.2610382}.
  \DOIprefix\doi{10.1145/2600428.2610382}.
\bibitem[{Tintarev and Masthoff(2007)}]{DBLP:conf/icde/TintarevM07}
\bibinfo{author}{N.~Tintarev}, \bibinfo{author}{J.~Masthoff},
\newblock \bibinfo{title}{A survey of explanations in recommender systems},
\newblock in: \bibinfo{booktitle}{{ICDE} Workshops}, \bibinfo{publisher}{{IEEE}
  Computer Society}, \bibinfo{year}{2007}, pp. \bibinfo{pages}{801--810}.
\bibitem[{Koren et~al.(2009)Koren, Bell, and
  Volinsky}]{DBLP:journals/computer/KorenBV09}
\bibinfo{author}{Y.~Koren}, \bibinfo{author}{R.~M. Bell},
  \bibinfo{author}{C.~Volinsky},
\newblock \bibinfo{title}{Matrix factorization techniques for recommender
  systems},
\newblock \bibinfo{journal}{Computer} \bibinfo{volume}{42}
  (\bibinfo{year}{2009}) \bibinfo{pages}{30--37}. \URLprefix
  \url{https://doi.org/10.1109/MC.2009.263}.
  \DOIprefix\doi{10.1109/MC.2009.263}.
\bibitem[{Tsukuda and Goto(2019)}]{DBLP:conf/recsys/TsukudaG19}
\bibinfo{author}{K.~Tsukuda}, \bibinfo{author}{M.~Goto},
\newblock \bibinfo{title}{Dualdiv: diversifying items and explanation styles in
  explainable hybrid recommendation},
\newblock in: \bibinfo{editor}{T.~Bogers}, \bibinfo{editor}{A.~Said},
  \bibinfo{editor}{P.~Brusilovsky}, \bibinfo{editor}{D.~Tikk} (Eds.),
  \bibinfo{booktitle}{Proceedings of the 13th {ACM} Conference on Recommender
  Systems, RecSys 2019, Copenhagen, Denmark, September 16-20, 2019},
  \bibinfo{publisher}{{ACM}}, \bibinfo{year}{2019}, pp.
  \bibinfo{pages}{398--402}. \URLprefix
  \url{https://doi.org/10.1145/3298689.3347063}.
  \DOIprefix\doi{10.1145/3298689.3347063}.
\bibitem[{Chen et~al.(2019)Chen, Chen, Xu, Zhang, Cao, Qin, and
  Zha}]{DBLP:conf/sigir/ChenCXZ0QZ19}
\bibinfo{author}{X.~Chen}, \bibinfo{author}{H.~Chen}, \bibinfo{author}{H.~Xu},
  \bibinfo{author}{Y.~Zhang}, \bibinfo{author}{Y.~Cao},
  \bibinfo{author}{Z.~Qin}, \bibinfo{author}{H.~Zha},
\newblock \bibinfo{title}{Personalized fashion recommendation with visual
  explanations based on multimodal attention network: Towards visually
  explainable recommendation},
\newblock in: \bibinfo{editor}{B.~Piwowarski}, \bibinfo{editor}{M.~Chevalier},
  \bibinfo{editor}{{\'{E}}.~Gaussier}, \bibinfo{editor}{Y.~Maarek},
  \bibinfo{editor}{J.~Nie}, \bibinfo{editor}{F.~Scholer} (Eds.),
  \bibinfo{booktitle}{Proceedings of the 42nd International {ACM} {SIGIR}
  Conference on Research and Development in Information Retrieval, {SIGIR}
  2019, Paris, France, July 21-25, 2019}, \bibinfo{publisher}{{ACM}},
  \bibinfo{year}{2019}, pp. \bibinfo{pages}{765--774}. \URLprefix
  \url{https://doi.org/10.1145/3331184.3331254}.
  \DOIprefix\doi{10.1145/3331184.3331254}.
\bibitem[{Cornacchia et~al.(2021)Cornacchia, Donini, Narducci, Pomo, and
  Ragone}]{DBLP:conf/caise/CornacchiaDNPR21}
\bibinfo{author}{G.~Cornacchia}, \bibinfo{author}{F.~M. Donini},
  \bibinfo{author}{F.~Narducci}, \bibinfo{author}{C.~Pomo},
  \bibinfo{author}{A.~Ragone},
\newblock \bibinfo{title}{Explanation in multi-stakeholder recommendation for
  enterprise decision support systems},
\newblock in: \bibinfo{editor}{A.~Polyvyanyy},
  \bibinfo{editor}{S.~Rinderle{-}Ma} (Eds.), \bibinfo{booktitle}{Advanced
  Information Systems Engineering Workshops - CAiSE 2021 International
  Workshops, Melbourne, VIC, Australia, June 28 - July 2, 2021, Proceedings},
  volume \bibinfo{volume}{423} of \textit{\bibinfo{series}{Lecture Notes in
  Business Information Processing}}, \bibinfo{publisher}{Springer},
  \bibinfo{year}{2021}, pp. \bibinfo{pages}{39--47}. \URLprefix
  \url{https://doi.org/10.1007/978-3-030-79022-6\_4}.
  \DOIprefix\doi{10.1007/978-3-030-79022-6\_4}.
\bibitem[{Lipton(2018)}]{DBLP:journals/cacm/Lipton18}
\bibinfo{author}{Z.~C. Lipton},
\newblock \bibinfo{title}{The mythos of model interpretability},
\newblock \bibinfo{journal}{Commun. {ACM}} \bibinfo{volume}{61}
  (\bibinfo{year}{2018}) \bibinfo{pages}{36--43}. \URLprefix
  \url{https://doi.org/10.1145/3233231}. \DOIprefix\doi{10.1145/3233231}.
\bibitem[{Wang et~al.(2018)Wang, Chen, Yang, Wu, Wu, and
  Xie}]{DBLP:conf/icdm/WangCYWW018}
\bibinfo{author}{X.~Wang}, \bibinfo{author}{Y.~Chen},
  \bibinfo{author}{J.~Yang}, \bibinfo{author}{L.~Wu}, \bibinfo{author}{Z.~Wu},
  \bibinfo{author}{X.~Xie},
\newblock \bibinfo{title}{A reinforcement learning framework for explainable
  recommendation},
\newblock in: \bibinfo{booktitle}{{IEEE} International Conference on Data
  Mining, {ICDM} 2018, Singapore, November 17-20, 2018},
  \bibinfo{publisher}{{IEEE} Computer Society}, \bibinfo{year}{2018}, pp.
  \bibinfo{pages}{587--596}. \URLprefix
  \url{https://doi.org/10.1109/ICDM.2018.00074}.
  \DOIprefix\doi{10.1109/ICDM.2018.00074}.
\bibitem[{Peake and Wang(2018)}]{DBLP:conf/kdd/PeakeW18}
\bibinfo{author}{G.~Peake}, \bibinfo{author}{J.~Wang},
\newblock \bibinfo{title}{Explanation mining: Post hoc interpretability of
  latent factor models for recommendation systems},
\newblock in: \bibinfo{editor}{Y.~Guo}, \bibinfo{editor}{F.~Farooq} (Eds.),
  \bibinfo{booktitle}{Proceedings of the 24th {ACM} {SIGKDD} International
  Conference on Knowledge Discovery {\&} Data Mining, {KDD} 2018, London, UK,
  August 19-23, 2018}, \bibinfo{publisher}{{ACM}}, \bibinfo{year}{2018}, pp.
  \bibinfo{pages}{2060--2069}. \URLprefix
  \url{https://doi.org/10.1145/3219819.3220072}.
  \DOIprefix\doi{10.1145/3219819.3220072}.
\bibitem[{Tao et~al.(2019)Tao, Jia, Wang, and Wang}]{DBLP:conf/sigir/TaoJWW19}
\bibinfo{author}{Y.~Tao}, \bibinfo{author}{Y.~Jia}, \bibinfo{author}{N.~Wang},
  \bibinfo{author}{H.~Wang},
\newblock \bibinfo{title}{The fact: Taming latent factor models for
  explainability with factorization trees},
\newblock in: \bibinfo{editor}{B.~Piwowarski}, \bibinfo{editor}{M.~Chevalier},
  \bibinfo{editor}{{\'{E}}.~Gaussier}, \bibinfo{editor}{Y.~Maarek},
  \bibinfo{editor}{J.~Nie}, \bibinfo{editor}{F.~Scholer} (Eds.),
  \bibinfo{booktitle}{Proceedings of the 42nd International {ACM} {SIGIR}
  Conference on Research and Development in Information Retrieval, {SIGIR}
  2019, Paris, France, July 21-25, 2019}, \bibinfo{publisher}{{ACM}},
  \bibinfo{year}{2019}, pp. \bibinfo{pages}{295--304}. \URLprefix
  \url{https://doi.org/10.1145/3331184.3331244}.
  \DOIprefix\doi{10.1145/3331184.3331244}.
\bibitem[{Gao et~al.(2019)Gao, Wang, Wang, and Xie}]{DBLP:conf/aaai/GaoWW019}
\bibinfo{author}{J.~Gao}, \bibinfo{author}{X.~Wang}, \bibinfo{author}{Y.~Wang},
  \bibinfo{author}{X.~Xie},
\newblock \bibinfo{title}{Explainable recommendation through attentive
  multi-view learning},
\newblock in: \bibinfo{booktitle}{The Thirty-Third {AAAI} Conference on
  Artificial Intelligence, {AAAI} 2019, The Thirty-First Innovative
  Applications of Artificial Intelligence Conference, {IAAI} 2019, The Ninth
  {AAAI} Symposium on Educational Advances in Artificial Intelligence, {EAAI}
  2019, Honolulu, Hawaii, USA, January 27 - February 1, 2019},
  \bibinfo{publisher}{{AAAI} Press}, \bibinfo{year}{2019}, pp.
  \bibinfo{pages}{3622--3629}. \URLprefix
  \url{https://doi.org/10.1609/aaai.v33i01.33013622}.
  \DOIprefix\doi{10.1609/aaai.v33i01.33013622}.
\bibitem[{Fusco et~al.(2019)Fusco, Vlachos, Vasileiadis, Wardatzky, and
  Schneider}]{DBLP:conf/ijcai/FuscoVVWS19}
\bibinfo{author}{F.~Fusco}, \bibinfo{author}{M.~Vlachos},
  \bibinfo{author}{V.~Vasileiadis}, \bibinfo{author}{K.~Wardatzky},
  \bibinfo{author}{J.~Schneider},
\newblock \bibinfo{title}{Reconet: An interpretable neural architecture for
  recommender systems},
\newblock in: \bibinfo{editor}{S.~Kraus} (Ed.), \bibinfo{booktitle}{Proceedings
  of the Twenty-Eighth International Joint Conference on Artificial
  Intelligence, {IJCAI} 2019, Macao, China, August 10-16, 2019},
  \bibinfo{publisher}{ijcai.org}, \bibinfo{year}{2019}, pp.
  \bibinfo{pages}{2343--2349}. \URLprefix
  \url{https://doi.org/10.24963/ijcai.2019/325}.
  \DOIprefix\doi{10.24963/ijcai.2019/325}.
\bibitem[{Tsang et~al.(2020)Tsang, Cheng, Liu, Feng, Zhou, and
  Liu}]{DBLP:conf/iclr/TsangCLFZL20}
\bibinfo{author}{M.~Tsang}, \bibinfo{author}{D.~Cheng},
  \bibinfo{author}{H.~Liu}, \bibinfo{author}{X.~Feng},
  \bibinfo{author}{E.~Zhou}, \bibinfo{author}{Y.~Liu},
\newblock \bibinfo{title}{Feature interaction interpretability: {A} case for
  explaining ad-recommendation systems via neural interaction detection},
\newblock in: \bibinfo{booktitle}{8th International Conference on Learning
  Representations, {ICLR} 2020, Addis Ababa, Ethiopia, April 26-30, 2020},
  \bibinfo{publisher}{OpenReview.net}, \bibinfo{year}{2020}. \URLprefix
  \url{https://openreview.net/forum?id=BkgnhTEtDS}.
\bibitem[{Strumbelj and Kononenko(2014)}]{DBLP:journals/kais/StrumbeljK14}
\bibinfo{author}{E.~Strumbelj}, \bibinfo{author}{I.~Kononenko},
\newblock \bibinfo{title}{Explaining prediction models and individual
  predictions with feature contributions},
\newblock \bibinfo{journal}{Knowl. Inf. Syst.} \bibinfo{volume}{41}
  (\bibinfo{year}{2014}) \bibinfo{pages}{647--665}. \URLprefix
  \url{https://doi.org/10.1007/s10115-013-0679-x}.
  \DOIprefix\doi{10.1007/s10115-013-0679-x}.
\bibitem[{Alvarez{-}Melis and
  Jaakkola(2018{\natexlab{a}})}]{DBLP:conf/nips/Alvarez-MelisJ18}
\bibinfo{author}{D.~Alvarez{-}Melis}, \bibinfo{author}{T.~S. Jaakkola},
\newblock \bibinfo{title}{Towards robust interpretability with self-explaining
  neural networks},
\newblock in: \bibinfo{editor}{S.~Bengio}, \bibinfo{editor}{H.~M. Wallach},
  \bibinfo{editor}{H.~Larochelle}, \bibinfo{editor}{K.~Grauman},
  \bibinfo{editor}{N.~Cesa{-}Bianchi}, \bibinfo{editor}{R.~Garnett} (Eds.),
  \bibinfo{booktitle}{Advances in Neural Information Processing Systems 31:
  Annual Conference on Neural Information Processing Systems 2018, NeurIPS
  2018, December 3-8, 2018, Montr{\'{e}}al, Canada},
  \bibinfo{year}{2018}{\natexlab{a}}, pp. \bibinfo{pages}{7786--7795}.
  \URLprefix
  \url{https://proceedings.neurips.cc/paper/2018/hash/3e9f0fc9b2f89e043bc6233994dfcf76-Abstract.html}.
\bibitem[{Alvarez{-}Melis and
  Jaakkola(2018{\natexlab{b}})}]{DBLP:journals/corr/abs-1806-08049}
\bibinfo{author}{D.~Alvarez{-}Melis}, \bibinfo{author}{T.~S. Jaakkola},
\newblock \bibinfo{title}{On the robustness of interpretability methods},
\newblock \bibinfo{journal}{CoRR} \bibinfo{volume}{abs/1806.08049}
  (\bibinfo{year}{2018}{\natexlab{b}}). \URLprefix
  \url{http://arxiv.org/abs/1806.08049}.
  \href{http://arxiv.org/abs/1806.08049}{{\tt arXiv:1806.08049}}.
\bibitem[{Saito et~al.(2020)Saito, Chua, Capel, and
  Hu}]{DBLP:journals/corr/abs-2006-12302}
\bibinfo{author}{S.~Saito}, \bibinfo{author}{E.~Chua},
  \bibinfo{author}{N.~Capel}, \bibinfo{author}{R.~Hu},
\newblock \bibinfo{title}{Improving {LIME} robustness with smarter locality
  sampling},
\newblock \bibinfo{journal}{CoRR} \bibinfo{volume}{abs/2006.12302}
  (\bibinfo{year}{2020}). \URLprefix \url{https://arxiv.org/abs/2006.12302}.
  \href{http://arxiv.org/abs/2006.12302}{{\tt arXiv:2006.12302}}.
\bibitem[{Slack et~al.(2020)Slack, Hilgard, Jia, Singh, and
  Lakkaraju}]{DBLP:conf/aies/SlackHJSL20}
\bibinfo{author}{D.~Slack}, \bibinfo{author}{S.~Hilgard},
  \bibinfo{author}{E.~Jia}, \bibinfo{author}{S.~Singh},
  \bibinfo{author}{H.~Lakkaraju},
\newblock \bibinfo{title}{Fooling {LIME} and {SHAP:} adversarial attacks on
  post hoc explanation methods},
\newblock in: \bibinfo{editor}{A.~N. Markham}, \bibinfo{editor}{J.~Powles},
  \bibinfo{editor}{T.~Walsh}, \bibinfo{editor}{A.~L. Washington} (Eds.),
  \bibinfo{booktitle}{{AIES} '20: {AAAI/ACM} Conference on AI, Ethics, and
  Society, New York, NY, USA, February 7-8, 2020}, \bibinfo{publisher}{{ACM}},
  \bibinfo{year}{2020}, pp. \bibinfo{pages}{180--186}. \URLprefix
  \url{https://doi.org/10.1145/3375627.3375830}.
  \DOIprefix\doi{10.1145/3375627.3375830}.
\bibitem[{Harper and Konstan(2016)}]{DBLP:journals/tiis/HarperK16}
\bibinfo{author}{F.~M. Harper}, \bibinfo{author}{J.~A. Konstan},
\newblock \bibinfo{title}{The movielens datasets: History and context},
\newblock \bibinfo{journal}{{ACM} Trans. Interact. Intell. Syst.}
  \bibinfo{volume}{5} (\bibinfo{year}{2016}) \bibinfo{pages}{19:1--19:19}.
  \URLprefix \url{https://doi.org/10.1145/2827872}.
  \DOIprefix\doi{10.1145/2827872}.
\bibitem[{Anelli et~al.(2021)Anelli, Bellog{\'{\i}}n, Ferrara, Malitesta,
  Merra, Pomo, Donini, and Noia}]{DBLP:conf/sigir/AnelliBFMMPDN21}
\bibinfo{author}{V.~W. Anelli}, \bibinfo{author}{A.~Bellog{\'{\i}}n},
  \bibinfo{author}{A.~Ferrara}, \bibinfo{author}{D.~Malitesta},
  \bibinfo{author}{F.~A. Merra}, \bibinfo{author}{C.~Pomo},
  \bibinfo{author}{F.~M. Donini}, \bibinfo{author}{T.~D. Noia},
\newblock \bibinfo{title}{Elliot: {A} comprehensive and rigorous framework for
  reproducible recommender systems evaluation},
\newblock in: \bibinfo{editor}{F.~Diaz}, \bibinfo{editor}{C.~Shah},
  \bibinfo{editor}{T.~Suel}, \bibinfo{editor}{P.~Castells},
  \bibinfo{editor}{R.~Jones}, \bibinfo{editor}{T.~Sakai} (Eds.),
  \bibinfo{booktitle}{{SIGIR} '21: The 44th International {ACM} {SIGIR}
  Conference on Research and Development in Information Retrieval, Virtual
  Event, Canada, July 11-15, 2021}, \bibinfo{publisher}{{ACM}},
  \bibinfo{year}{2021}, pp. \bibinfo{pages}{2405--2414}. \URLprefix
  \url{https://doi.org/10.1145/3404835.3463245}.
  \DOIprefix\doi{10.1145/3404835.3463245}.
\bibitem[{Anelli et~al.(2019)Anelli, Noia, Sciascio, Pomo, and
  Ragone}]{DBLP:conf/recsys/AnelliNSPR19}
\bibinfo{author}{V.~W. Anelli}, \bibinfo{author}{T.~D. Noia},
  \bibinfo{author}{E.~D. Sciascio}, \bibinfo{author}{C.~Pomo},
  \bibinfo{author}{A.~Ragone},
\newblock \bibinfo{title}{On the discriminative power of hyper-parameters in
  cross-validation and how to choose them},
\newblock in: \bibinfo{editor}{T.~Bogers}, \bibinfo{editor}{A.~Said},
  \bibinfo{editor}{P.~Brusilovsky}, \bibinfo{editor}{D.~Tikk} (Eds.),
  \bibinfo{booktitle}{Proceedings of the 13th {ACM} Conference on Recommender
  Systems, RecSys 2019, Copenhagen, Denmark, September 16-20, 2019},
  \bibinfo{publisher}{{ACM}}, \bibinfo{year}{2019}, pp.
  \bibinfo{pages}{447--451}. \URLprefix
  \url{https://doi.org/10.1145/3298689.3347010}.
  \DOIprefix\doi{10.1145/3298689.3347010}.
\bibitem[{Krichene and Rendle(2020)}]{DBLP:conf/kdd/KricheneR20}
\bibinfo{author}{W.~Krichene}, \bibinfo{author}{S.~Rendle},
\newblock \bibinfo{title}{On sampled metrics for item recommendation},
\newblock in: \bibinfo{editor}{R.~Gupta}, \bibinfo{editor}{Y.~Liu},
  \bibinfo{editor}{J.~Tang}, \bibinfo{editor}{B.~A. Prakash} (Eds.),
  \bibinfo{booktitle}{{KDD} '20: The 26th {ACM} {SIGKDD} Conference on
  Knowledge Discovery and Data Mining, Virtual Event, CA, USA, August 23-27,
  2020}, \bibinfo{publisher}{{ACM}}, \bibinfo{year}{2020}, pp.
  \bibinfo{pages}{1748--1757}. \URLprefix
  \url{https://doi.org/10.1145/3394486.3403226}.
  \DOIprefix\doi{10.1145/3394486.3403226}.

\end{thebibliography}

\end{document}